\documentclass[superscriptaddress, reprint, amsmath, amssymb, aps, pra,
floatfix]{revtex4-2}

\usepackage{dcolumn}
\usepackage{amsfonts}
\usepackage{amsmath}
\usepackage{amsthm}
\usepackage{graphics}
\usepackage{graphicx}
\usepackage{subfigure}
\usepackage{amssymb}
\usepackage{graphics}
\usepackage{graphicx}
\usepackage{epstopdf}
\usepackage{color}
\usepackage{subfigure}
\usepackage{pgfplots,tikz}
\usepackage{url}
\usepackage{float}
\usepackage{soul}
\def\erfc{{\rm erfc}}

\begin{document}

\title{Vortex to Rotons Transition in Dipolar Bose-Einstein Condensates}

\author{Alberto Villois}
\affiliation{School of Engineering, Mathematics and Physics, University of East Anglia, Norwich Research Park, Norwich, NR4 7TJ, United Kingdom}
\affiliation{Centre for Photonics and Quantum Science, University of East Anglia, Norwich Research Park, Norwich, NR4 7TJ, United Kingdom}
\affiliation{Dipartimento di Fisica, Universit{\`a} degli Studi di Torino, Via Pietro Giuria 1, 10125 Torino, Italy}

\author{Miguel Onorato}
\affiliation{Dipartimento di Fisica, Universit{\`a} degli Studi di Torino, Via Pietro Giuria 1, 10125 Torino, Italy}
\affiliation{INFN, Sezione di Torino, Via Pietro Giuria 1, 10125 Torino, Italy}
\pacs{}

\author{Davide Proment}
\affiliation{School of Engineering, Mathematics and Physics, University of East Anglia, Norwich Research Park, Norwich, NR4 7TJ, United Kingdom}
\affiliation{Centre for Photonics and Quantum Science, University of East Anglia, Norwich Research Park, Norwich, NR4 7TJ, United Kingdom}
\affiliation{ExtreMe Matter Institute EMMI, GSI Helmholtzzentrum fuer Schwerionenforschung, Planckstrasse 1, 64291 Darmstadt, Germany}

\begin{abstract}
Dipolar Bose-Einstein condensates (dBECs) exhibit a plethora of physics phenomena, from supersolidity to roton-like minimum in the elementary excitation spectrum.
In this work we first demonstrate the existence of axis-symmetric solitary waves in (quasi-)two-dimensional dBECs: these localised excitations are characterised by quantised vortex dipoles that continuously transit to vortex-free density depletions.
We then show how the presence of the roton minimum fundamentally alters the fate of such solutions when approaching the Landau's critical speed: when propagating along the polarisation direction where the roton minimum occurs, the solitary wave transits into roton excitations rather than into phonons as for standard contact-interaction BECs.
This finding suggests that Feynman’s hypothesis, conjectured for 3D superfluid liquid helium regarding the creation of rotons as fading vortex excitations, is valid in the context of 2D dBECs.

\end{abstract}
\maketitle

{\it Introduction}.
Solitary waves stand out as a primary feature in nonlinear physics, finding application across diverse fields including fluid dynamics, optics, and plasma physics, while offering profound insights into the fundamental behaviour of complex systems. 
These self-sustaining wave patterns defy the typical dispersion expected in wave motion, maintaining their shape and velocity as they propagate through a medium. 
Their resilience against dispersion stems from a delicate balance between nonlinear and dispersive effects, making them stable entities within nonlinear systems.

Solitary waves characterised by the presence of vorticity have particular significance within the field of quantum fluids, first discovered in the context of superfluid liquid helium \cite{donnelly1991quantized}.
Due to the topological nature of quantised vorticity, such structures play a fundamental role in the study of vortex dynamics and quantum turbulence \cite{PhysRevLett.100.245301}. 
Notable examples are the vortex ring cascade hypothesis in three dimensions superfluids \cite{FEYNMAN195517, Nemirovskii2014} and the Berezinskii-Kosterlitz-Thouless transition in two spatial dimensions \cite{hadzibabic2011two}. Moreover, the study of such solitary waves has significantly contributed to the understanding of the nature of the superfluid. 
Specifically, the nucleation of solitary waves with quantised vorticity is believed, alongside nucleation of rotons, to be one of the possible mechanisms for breaking superfluidity in liquid helium, particularly in the low-pressure regime \cite{Nancolas1985, doi:10.1098/rsta.1985.0041}.

The relationship between roton excitations and solitary waves has long been a subject of fascination. Particularly, Feynman postulated the origin of drag in a superfluid as the generation of roton excitations from a vanishing vortex ring \cite{FEYNMAN195517}, sparking numerous studies aimed at determining whether the energy and momentum of vortex rings were comparable to roton excitations.
However, this hypothesis was disproved in the seminal work of Jones \& Roberts \cite{Jones:1982aa}:
With a weakly interacting model aimed to capture the key features of superfluid liquid helium, they conducted an investigation into the existence of a family of axisymmetric solitary waves.
These waves exhibit quantised vorticity at low speeds, taking the form of vortex rings in three dimensions and vortex dipoles in two dimensions. As their speed increases, they transition into vortex-free density dips, reaching speeds comparable to speed of phonons, the fastest excitations in the system before superfluidity breaks down \cite{pitaevskii2016bose}.

Over the years, numerous experimental advancements have facilitated the generation of various systems exhibiting superfluid behaviour, ranging from single and multicomponent Bose-Einstein condensates (BECs) to spinor and Fermi gases. One of the most notable achievements was the development of BECs characterised by long-range dipole-dipole interactions, known as dipolar BECs (dBECs) \cite{Chomaz_2023, Bigagli2024}. 
When one or more spatial dimensions are strongly confined, these condensates have elementary excitations which display a dispersion relation that, akin to superfluid liquid helium \cite{, PhysRevB.103.104516}, features a roton minimum \cite{PhysRevA.73.031602,PhysRevLett.90.250403, PhysRevLett.90.110402, PhysRevLett.106.065301}. 
Such elementary excitations have recently been experimentally measured in dBECs \cite{, Chomaz2018, PhysRevLett.122.183401} and enabled the generation of a new supersolid state of matter \cite{PhysRevLett.122.130405,PhysRevX.9.011051,PhysRevX.9.021012}. 
Unlike liquid helium, cold gases allow for the direct visualisation of small-scale vortex structures, as demonstrated recently \cite{PhysRevLett.119.150403, PhysRevLett.115.170402, Klaus2022}. 
Moreover, they can be accurately described by mean-field models based on the Gross-Pitaevskii equation. 
These properties make such systems particularly compelling for exploring the interplay between vortical structures and rotons. 

In this Letter, we investigate the existence of the two-dimensional Jones-Roberts solitary wave family in a dBEC, from solutions characterised by topological defects, to vortex-free density depletions, also known as Jones-Roberts (JR) solitons~\cite{PhysRevLett.119.150403}. 
The explicit connection between solitary waves and roton excitations is also demonstrated for the first time, providing support for Feynman’s hypothesis regarding the creation of rotons as fading vortex excitations.

\begin{figure*}
    \centering
    \includegraphics[width=\textwidth]{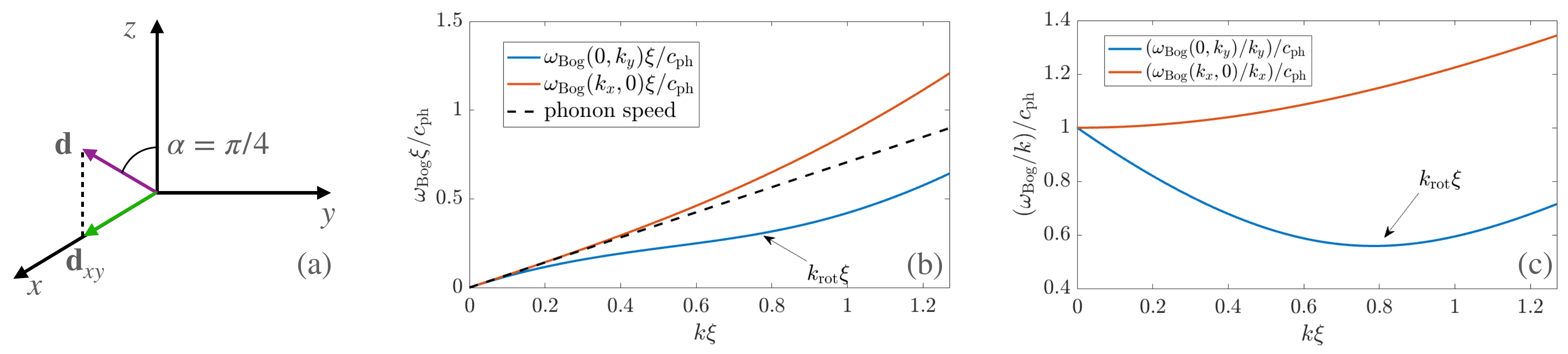}
    \caption{In all plots we take $ \alpha=\pi/4 $ and $ \beta=0.9 $.
(a) Sketch of the orientation of the dipole moment. 
(b) Bogoliubov dispersion relation along the $ k_x $ (red) and $ k_y $-axis (blue). The phonon limit is shown with a dashed line.
(c) Phase speed of the Bogoliubov excitations along the $ k_x $ (red) and $ k_y $-axis (blue). 
The minimum value of each curve in panel (c) indicates the limiting speed of the localised solitary wave, that is the minimum speed at which eq.~(7) is broken along the $ x $ (red) and $ y $ (blue) direction, respectively.
}
    \label{fig:1}
\end{figure*}

{\it Theoretical model.}
In  the limit where the condensate is strongly confined along the $ z $-direction 
with a characteristic confining length scale $ l_z $, 
an excellent (quasi-)two-dimensional model with no confinement in the $ x $-$ y $ plane is given by the following Gross-Pitaevskii equation
\begin{equation}
    i\hbar\frac{\partial\psi}{\partial t} = - \frac{\hbar^2}{2m}\nabla^2\psi +g_{\rm c}|\psi|^2\psi + g_{\rm d}\Phi\psi \, .
    \label{eq:GP2D}
\end{equation}
Here, $ \psi(x, y, t) $ is the two-dimensional order parameter, $\nabla^2=\partial_{xx}+\partial_{yy}$, $ g_{\rm c} = 2\sqrt{2\pi}\hbar^2 a_{\rm s}/(m l_z) $ is the contact (local) interaction coefficient, $ m $ and $ a_{\rm s} $ are the mass and the s-wave scattering length of the boson, respectively,
and $ g_{\rm d} = |{\bf d}|^2/\sqrt{2\pi} l_z $ is the dipole-dipole (nonlocal) interaction coefficient where $ \bf d $ is the dipole moment.
The nonlocal dipolar interaction operator is defined as
\begin{equation}
    \Phi=\frac{4\pi}{3} \mathcal{F}_{xy}^{-1}\left[ \mathcal{F}_{xy}\left(|\psi|^2\right) F\left({\bf q}\right) \right]
\end{equation}
where $ \mathcal{F}_{xy}(\cdot) $ is the spatial two-dimensional direct Fourier transform operator 
\footnote{Here we use the normalisation and sign convention $ \mathcal{F}_{xy}(\cdot) = \int (\cdot) e^{-i(k_x x + k_y y)} dxdy $ for the direct Fourier Transform in two spatial dimensions, resulting in the inverse $ \mathcal{F}_{xy}^{-1}(\cdot) = 1/(2\pi) \int (\cdot) e^{i(k_x x + k_y y)} dk_xdk_y $.}
and ${\bf q}= {\bf k}l_z/\sqrt{2} $ is the wave vector on the $x$-$y$ plane.
 Without any loss of generality, we consider the dipole moment whose projection onto the $x$-$y$ plane is aligned with the positive $x$-axis and define $ \alpha $ being the angle between $ {\bf d} $ and the positive $ z $-axis, see Fig.~\ref{fig:1}~(a).
With this choice the function $F(\cdot)$, that represents the Fourier transform of the two-dimensional dipole-dipole interaction \cite{PhysRevLett.106.065301, Mulkerin_2014}, results in
\begin{equation}
    F({\bf q})=\cos^2\alpha F_\perp(q) + \sin^2\alpha F_\parallel(q, q_x) \, ,
    \label{eq:F}
\end{equation} 
where $ q = |{\bf q}| $, $ q_x $ being the $ x $-component of $ {\bf q} $, and 
\begin{eqnarray*}
&&F_\perp(q)=2 - 3\sqrt{\pi} q e^{q^2} \erfc(q) \\
&&F_\parallel(q, q_x)=-1 + 3\sqrt{\pi} \frac{q_x^2}{q}e^{q^2} \erfc(q) \, .
\end{eqnarray*}
By assuming the homogeneous solution in the form
\begin{equation}
    \psi(t) = \sqrt{n} e^{-i\frac{\mu t}{\hbar}}
    \label{eq:homoSol} 
\end{equation} 
where $ n $ is the average particle number density per unit of volume, one finds the chemical potential 
\begin{equation}
    \mu = g_{\rm c} n \left[ 1 + \left(3\cos^2\alpha - 1\right)\frac{4\pi}{3}\beta\right] \, , \quad \text{with $ \beta=\frac{g_{\rm d}}{g_{\rm c}} $ \footnote{note that in the dipolar gases literature $ \beta $ is often labelled as $ e_{dd} $.},}
    \label{eq:mu}
\end{equation}
and its infinitesimal wave-like perturbations of wavelength $ {\bf k}=(k_x, k_y) $, and angular frequency $ \omega_{\rm Bog} $ to satisfy the Bogoliubov dispersion relation \cite{PhysRevLett.106.065301}

\begin{equation}
    \omega_{\rm Bog}({\bf k}) = \pm\frac{|{\bf k}|}{\sqrt{m}} \sqrt{\frac{\hbar^2 |{\bf k}|^2}{4m} + g_{\rm c}n\left[1+\beta\frac{4\pi}{3}F\left(\frac{l_z {\bf k}}{\sqrt{2}}\right)\right]} \, .
\end{equation}
Finally, by introducing the healing length, $\xi = \hbar/\sqrt{2 m \mu} $, one finds the phonon speed (long wave perturbation) as $ c_{\rm ph}=\sqrt{\mu/m} $.
One notices that, when $ \alpha \neq 0 $, Eq.~(\ref{eq:F}) is anisotropic in $ {\bf k} $, hence the Bogoliubov dispersion.
Figure~\ref{fig:1}~(b) illustrates the two dispersion relation branches along the $ k_x $ and $ k_y $ directions, choosing $ \alpha=\pi/4 $ and $ \beta=0.9 $.
For such values, used for all the results presented in this Letter, the dispersion relation along the $ k_y $ direction shows a point where the quantity $ \omega(k_y, 0)/k_y $ possesses a global minimum and corresponds to an analogue of the roton minimum measured in superfluid liquid helium, therefore called here {\it roton-like} minimum.
Note that this minimum is present for a large range of $ (\alpha, \beta) $, and our choice is achievable in experimental setups \cite{Chomaz_2023}.

\begin{figure*}
    \centering
    \includegraphics[width=\textwidth]{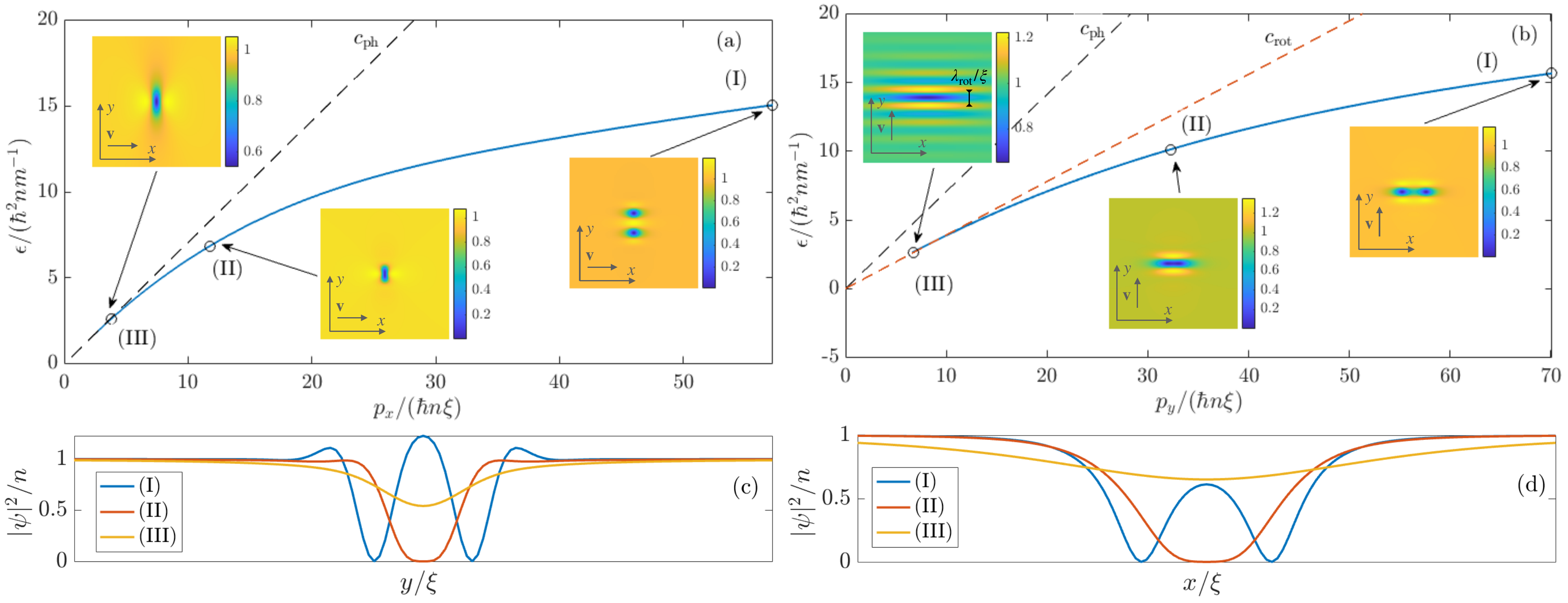}
    \caption{The blue line shows the energy-momentum plot of solitary wave family moving along the $ x $ and $ y $-direction, given in panels (a) and (b), respectively. 
Each point on the curve is a particular solitary wave whose speed is given by the slope of the tangent to the curve and its density profile $ |\psi|^2 $, rescaled by the average density $ n $ and zoomed around the localised density dip(s), is shown in the respective inset.
The point (II) indicate where the transition between
a vortex dipole to a JR soliton takes place, while point (III) shows an example of a density depletion.
The black and red dashed lines represent the phonon and roton speed, respectively.
Panel~(c) displays the density profiles along the $y$-axis at $x = L_x/2$ for the solutions labelled (I), (II), and (III) in panel~(a). 
Similarly, panel~(d) shows the density profiles along the $x$-axis at $y = L_y/2$ for the corresponding solutions presented in panel~(b).
}
    \label{fig:2}
\end{figure*}

Taking advantage of the anisotropy of the dispersion relation, in this Letter we investigate the existence of solitary wave solutions in two-dimensional dBECs, their dependence on the alignment between their propagation velocity and the dipole moment, and revisit Feynman's hypothesis, originally formulated in superfluid liquid helium, on the vortex-roton transition.
Jones \& Roberts have demonstrated the existence of axisymmetric solitary waves within the Gross-Pitaevskii equation characterised only by contact interactions \cite{Jones:1982aa}. Their results were reproduced experimentally in \cite{PhysRevLett.119.150403}.
In two dimensions, these solutions exhibit counter-rotating quantised vortices at low speeds that transition to density pulses free of topological defects at speeds comparable to the phonon speed.
It is pedagogical to identify first the range of possible speeds at which a solitary travelling wave
can propagate. 
To do so, one relies on the so-called band gap analysis introduced in \cite{Villois:19}.
 The idea behind the band gap analysis is the following: a localised (fully nonlinear) solitary wave excitation of the homogeneous solution~\eqref{eq:homoSol} moving at velocity $ {\bf v}=(v_x, v_y) $ can exist only if
\begin{equation}\label{Eq:Landau}
    {\bf v} \cdot {\bf k} \neq \omega_{\rm Bog}({\bf k}) \, .
\end{equation}
This is to prevent the resonant transfer of energy from the solitary wave to any delocalised Bogoliubov excitation of wave vector $ \mathbf{k} $.
By using this criterion, it is straightforward to identify the range of speed values that the solitary wave is allowed, as shown in Fig.~\ref{fig:1}~(c).
Note that Eq.~\eqref{Eq:Landau} is equivalent to Landau's criterion for the breakdown of superfluidity, demonstrating that the along the $ x $-direction superfluid excitations cannot move faster than the phonon speed, while along the $ y $-direction the limiting speed is given by the roton speed $ c_{\rm rot}={\rm min
}\left(\omega_{\rm Bog}(0, k_y)/k_y \right) $ \cite{PhysRevLett.121.030401}.

{\it Numerical methods.}
The solitary wave solution is computed numerically by recasting the Gross-Pitaevskii equation (\ref{eq:GP2D}) in the reference frame moving with velocity $ {\bf v} $ and seeking for a time-independent solution having the same chemical potential (\ref{eq:mu}) of the homogeneous solution (\ref{eq:homoSol}), see the SM for more details.
The system is spatially discretized on a regular lattice with $ 512^2 $ collocation points over a box with size $L_x/\xi = L_y/\xi=80\pi$. Periodic boundary conditions are considered in order to compute differential operators making use of Fourier spectral decomposition.
The time-independent solution is found via a Newton-Raphson method, and the JR branch is followed by slowly varying the solitary wave speed.
We also perform time evolutions of the time-dependent Gross-Pitaevskii equation (\ref{eq:GP2D}) to analyse qualitatively the stability of the solitary wave; the time operator is resolved using a standard RK4 method.
A thorough linear stability analysis is performed by looking at the spectral eigenvalues of the Bogoliubov-de Gennes (linearised) perturbations. 

{\it Numerical results.}
First, we investigate the family of solitary waves propagating along the direction of the dipole polarisation ($ x $-direction), that is choosing $ v_y = 0 $. 
Each point on the blue line in Fig.~\ref{fig:2}~(a) represents the energy-momentum coordinates of a solitary wave solution. 
The $ x $-component of the velocity of each solitary wave is determined by the slope of this curve, as given analytically by $ v_x = \partial\epsilon/\partial p_x |_{v_y=0} $, with $ \epsilon $ and $ p_x $ being the energy and $ x $-component of the linear momentum densities of the solitary wave, respectively (see the SM for the mathematical derivation). 
Density profiles zoomed about the localised solitary wave are shown in the insets.
Point (I) illustrates a vortex dipole (two topological defects in the argument of the order parameter, the order parameter goes to zero at the defect points), the transition between a solution with and without vorticity (no more topological defects, the order parameter still goes to zero) is indicated by point (II), point (III) shows a non-zero density depletion in the order parameter; Fig.~\ref{fig:2}~(c)) shows how the density profile along the $ y $-axis varies during the transition).
The vortex core density profiles in the dipole solution are elongated along the polarisation direction, in agreement with previous results~\cite{PhysRevA.109.063323,PhysRevLett.111.170402}. Here we show that such elongation is still present in the absence of vorticity, although it is lost in the limit for speeds approaching the phonon speed, see dashed black line in Fig.~\ref{fig:2}~(a), with a JR soliton resembling the typical profile of lump soliton solutions for the so-called Kadomtsev-Petviashvili equation~\cite{PhysRevLett.116.173901}.

We then explore the family of solitary waves moving perpendicularly to the direction of the dipole polarisation ($ y $-direction), that is choosing $ v_x=0 $, whose results are plotted in Fig.~\ref{fig:2}~(b).
Here, the $ y $-component of the velocity of each solitary wave is determined by the slope of this curve, resulting in $ v_y = \partial\epsilon/\partial p_y |_{v_x=0} $ where $ p_y $ is the $ y $-component of the linear momentum density of the solitary wave.
Again, density profiles zoomed about the localised waves are shown in the insets.
Similarly to the previous case, both vortex dipole solutions  and JR solitons have density profiles elongated along the polarisation direction, see point (I) and point (II) in Fig.~\ref{fig:2}~(b), respectively. Point (II) highlights the transition from vortex dipole to JR soliton, see Fig.~\ref{fig:2}~(d) for the density profile along the $ x $-axis. 
However, in this case the JR solitons cannot approach the phonon speed resembling lump soliton solutions, as Landau's criterion for superfluidity breaking along the $ y $-direction is now determined by the roton speed, shown by a dashed red line in Fig.~\ref{fig:2}~(b). 
Upon approaching the roton speed, a transition from localised JR solitons to delocalised roton excitations appearing as stripes of characteristic scale $ \lambda_{\rm rot} = 2\pi/k_{\rm rot} $ becomes qualitatively visible, see point (III) in Fig.~\ref{fig:2}~(b).
Figure~(1) in the SM shows the evolution of the density and phase fields covering a distance around $ 20 L_y $, demonstrating visually that the solitary wave associated with point (III) in Fig.~\ref{fig:2}~(b) is stable in time even when white noise is present.
A full Bogoliubov-de Gennes stability analysis, shown in the SM, has been carried out confirming the robustness of the solitary wave solutions with respect to noise.

\begin{figure}
    \centering
    \includegraphics[width=\columnwidth]{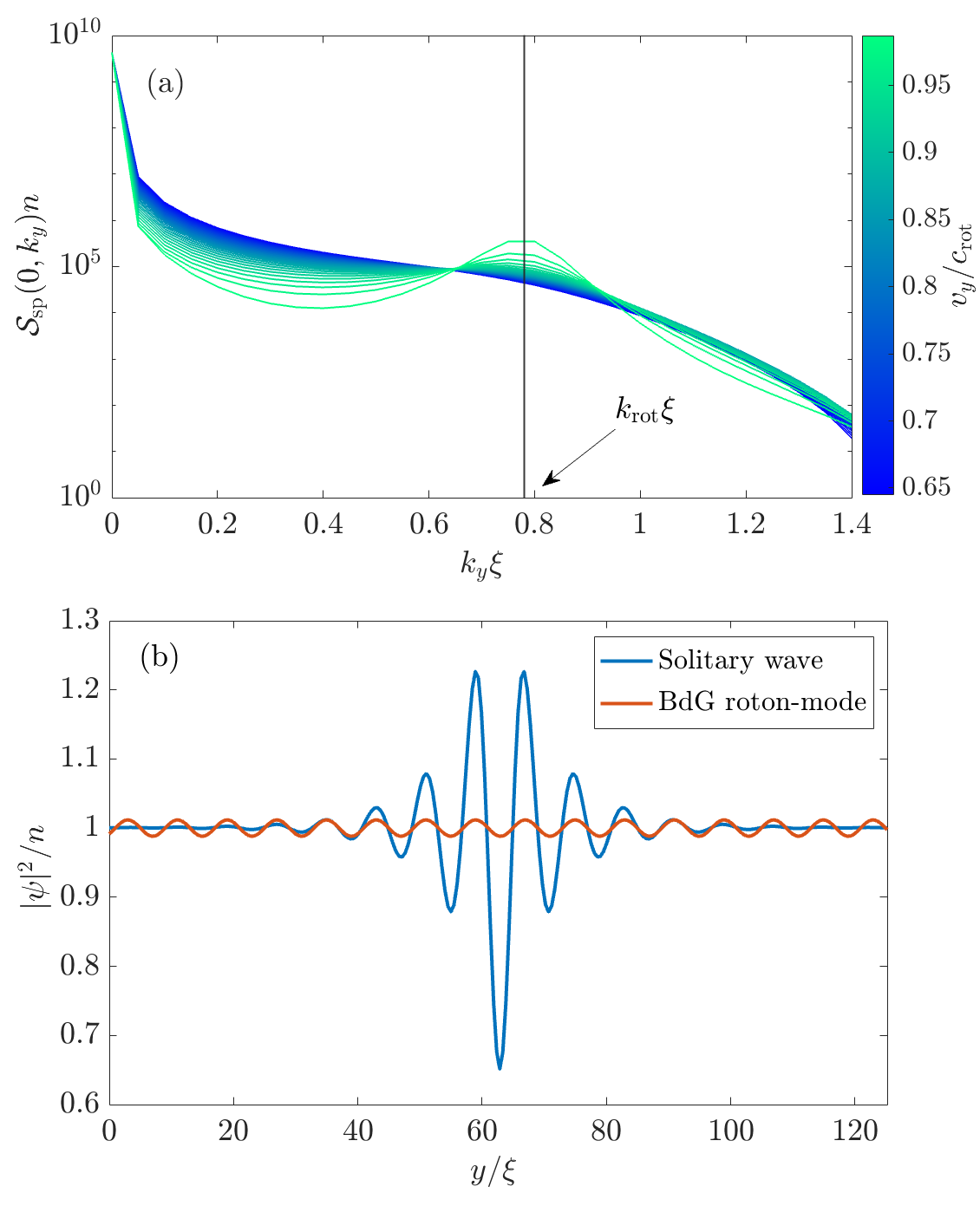}
    \caption{
    (a) Spatial spectra $ \mathcal{S}_{\rm sp}(k_x=0, k_y) $ of the solitary wave solutions moving along the $ y $-direction plotted for different $ v_y $ tending to the roton speed. (b) Profile of solitary wave associated to point (III) in Fig.~\ref{fig:2}~(b) compared with space profile or roton excitation from BdG analysis. 
    }
    \label{fig:3}
\end{figure}
We now analyse the spectral properties of the solitary wave branch presented in Fig.~\ref{fig:2}~(b).
Figure~\ref{fig:3}~(a) shows the collection of spatial spectra $ \mathcal{S}_{\rm sp}(k_x, k_y) = |\mathcal{F}_{xy}(\psi_{\rm SW})|^2 $ plotted along the $ k_y $-axis of the solitary waves found in Fig.~\ref{fig:2}~(b) for speeds, $v_y/c_{\rm rot} \ge 0.65 $.
It is evident that a peak around the roton-like minimum wave number $ k_{\rm rot} $ emerges and grows for growing speeds; this is a first indication that the travelling wave solution approaches the roton-like Bogoliubov excitation when its speed tends to the roton speed.

To further support the hypothesis of a continuous transition between a localised solution and a delocalised roton excitation, we plot in Fig.~\ref{fig:3}~(b) the comparison of the density profile along the \( y \)-axis of the localized travelling wave solution corresponding to point (III) in Fig.~\ref{fig:2}~(b) with the profile of a Bogoliubov mode associated with a roton excitation computed using the Bogoliubov-de Gennes approach, eq.~(22) in SM.
As the figure shows, the oscillatory behaviour of the two profiles tend to overlap across a significant portion of the domain.

Having demonstrated that a dBEC can host the JR solitary waves which continuously transit from a vortex dipole, to a density depletion and subsequently tends to a roton excitation, we now aim to investigate the likelihood of such a transition occurring dynamically, namely seeking the 2D analogue of Feynman’s hypothesis regarding the transition of a vortex ring into a roton. 
In Feynman’s original conjecture, such a transition occurs in turbulent settings as a result of scattering processes involving other vortices or thermal excitations.

In the simplest possible way, we consider the dissipative Gross-Pitaevskii equation \cite{Proukakis_2008}
\begin{equation}
(i - \gamma) \hbar\frac{\partial \psi}{\partial t} = - \frac{\hbar^2}{2m}\nabla^2\psi +g_{\rm c}|\psi|^2\psi + g_{\rm d}\Phi\psi + \mu\psi \, ,
\label{eq:dGPE}
\end{equation}
where \( \gamma \) is a phenomenological temperature-depended dissipation coefficient that models the loss of energy and momentum due to interactions with the thermal bath.

\begin{figure}
    \centering
    \includegraphics[width=\columnwidth]{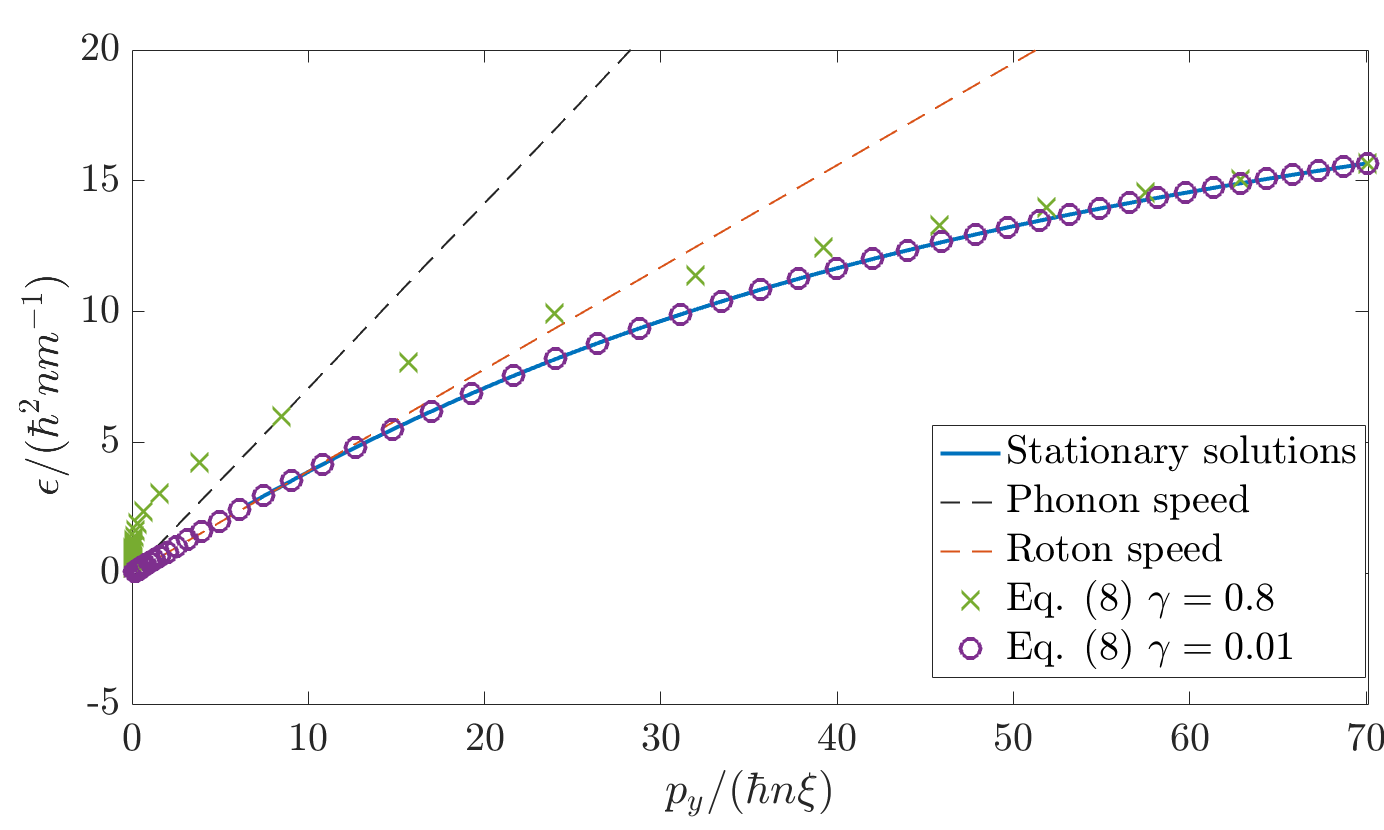}
    \caption{
   Energy vs. linear momentum along the $ y $-direction plot.
   The blue line shows the values of the JR solitary wave branch plotted in Fig.~\ref{fig:2}~(b).
   The circles ($ \gamma=0.01 $) and crosses ($ \gamma=0.8 $) show the values of the momentum and energy at successive time steps of the initial condition at point (I) in Fig.~\ref{fig:2}~(b) evolved in time according to eq.~(\ref{eq:dGPE}).
    }
    \label{fig:4}
\end{figure}

By numerically solving eq.~(\ref{eq:dGPE}) using the solitary wave at point (I) in Fig.~\ref{fig:2}~(b) as initial condition, we show in Fig.~\ref{fig:4} how values of the linear momentum along the $ y $-axis and energy of the initial condition decay, measuring them at subsequent time intervals and for two different dissipation coefficients, $ \gamma=0.8 $ and $ \gamma=0.01 $.
An animation of the time evolution of the density and phase fields, for $ \gamma=0.01 $, is also provided in the SM: this clearly shows that the initial vortex dipole solution dynamically transit first into a density depletion and then into a rotonic excitation. 
Interestingly, throughout this dynamical transition, the energy and momentum evolve following the same trend measured for the JR solitary branch of stationary solutions, especially when $ \gamma=0.01 \ll 1 $ (the overlap between the circles and the blue line in Fig.~\ref{fig:4} is striking).
When the dissipation parameter is larger, here $ \gamma=0.8 \sim 1 $, the full correspondence is broken, and the dynamical transition overestimates the JR solitary branch of stationary solutions.
Understanding the effects of strong dissipation may require alternative damping models~\cite{PhysRevA.110.053302, PhysRevLett.127.101601}, and this goes beyond the scope of the present work.

{\it Conclusions and outlook}.
In this Letter we demonstrated the existence of the Jones-Roberts solitary wave solutions in two-dimensional dipolar condensates, showing that the solution branch is formed by a vortex dipole that transits into a topologically-free density depletion when exceeding a critical speed.
The properties of the Jones-Robert branch strongly depend on the alignment between the solitary wave velocity and the dipole polarisation direction. 
When these are fully aligned, the maximum soliton speed approaches the phonon velocity, as in standard (non-dipolar) condensates. 
Conversely, when the motion is perpendicular to the polarisation direction, the limiting speed is determined by the threshold for roton-like excitations.
Our stability analyses, conducted both through direct simulations introducing small-amplitude noise and within the linearised Bogoliubov–de Gennes framework, confirm that these solitary waves are robust and should be experimentally realisable within the parameter range of state-of-the-art dipolar gas setups \cite{Kwon2021}.

Our results confirm, in dipolar gases, the validity of Feynman’s conjecture regarding the dynamical creation of rotons as fading vortex excitations. 
To this end, we investigated the behaviour of the Jones-Roberts solitons moving perpendicularly to the polarisation direction, that is when their speed is limited by the roton speed. 
As its velocity increases, the Jones-Roberts solitary wavs develop a spectral peak at the roton wavenumber and becomes less localised, gradually transforming into a delocalised roton-like Bogoliubov excitation.
Following Feynmann's idea, we examined the possibility of observing this transition dynamically.
By extending the model to include dissipative effects, we show that a transition from a vortex dipole to delocalised roton excitations is indeed possible, providing validation for Feynman’s interpretation of the generation of rotons.

Questions, such as 
how this transition affects vortex dynamics and turbulence in dipolar gases, whether a similar mechanism can be extended to superfluid liquid helium in three dimensions, the relation between the JR and Kadomtsev-Petviashvili solitons \cite{PhysRevLett.116.173901} in the Landau's critical speed limit, and if there exist systems where the Landau's critical speed is lower than the transition speed between vortex dipole and density depletion, remain open.

\begin{acknowledgments}
We are indebited to one of the anonymous Referees for recommending to explore the transition from JR solitons to roton excitations dynamically.
M.O. was funded by Progetti di Ricerca di Interesse Nazionale (Grants No. 2020X4T57A and No. 2022WKRYNL) and by the Simons Foundation (Grant No. 652354)
D.P. would like to thank the Isaac Newton Institute for Mathematical Sciences for support and hospitality during the programme Dispersive hydrodynamics: mathematics, simulation and experiments when the final part of the work on this paper was undertaken. 
This work was supported by EPSRC Grant Number EP/R014604/1.
This research was supported in part by the ExtreMe Matter Institute EMMI at the GSI Helmholtzzentrum fuer Schwerionenphysik, Darmstadt, Germany.
\end{acknowledgments}

\bibliographystyle{apsrev4-2}
\bibliography{./references}

\end{document}